\documentclass[aps,prd,preprint,groupedaddress]{revtex4-1}
\pdfoutput=1

\usepackage{graphicx}

\begin{document}


\title{
{\normalsize \noindent
MAN/HEP/2017/011
\hfill
\null\hskip 15mm IPPP/17/80
}\\
$\null$ \\
Strong coupling constant extraction from high-multiplicity $Z$+jets observables}


\author{Mark Johnson}
\email[]{mark.andrew.johnson@cern.ch}
\affiliation{Manchester University and the Cockcroft Institute}

\author{Daniel Ma\^{\i}tre}
\email[]{daniel.maitre@durham.ac.uk}
\affiliation{Institute for Particle Physics Phenomenology, Durham University}


\newcommand{\amz}{\alpha_S(M_Z)}
\newcommand{\BHS}{ {\sc BlackHat+Sherpa}}
\newcommand{\fastNLO}{{\sc fastNLO}}
\newcommand{\ntuple}{n-Tuple}
\newcommand{\ntuples}{n-Tuples}

\date{\today}

\begin{abstract}
We present a strong coupling constant extraction at Next-to-Leading Order QCD accuracy using ATLAS Z+2,3,4 jets data. This is the first extraction using processes with a dependency to high powers of the coupling constant. We obtain values of the strong coupling constant at the $Z$ mass compatible with the world average and with uncertainties commensurate with other NLO extractions at hadron colliders. Our most conservative result for the strong coupling constant is $\amz=
0.1178 ^{+0.0051 }_{ - 0.0043 }
$. 
\end{abstract}


\maketitle

\section{Introduction}

The strong coupling constant $\alpha_S$ is a physical parameter of QCD that cannot be predicted from first principles and has to be obtained from an experimental measurement. Values of $\alpha_S$ have been previously obtained by comparing experimental data from hadronic $\tau$ decays, deep inelastic scattering, heavy quarkonia decays or measurements from $e^+e^-$ and hadron colliders against theoretical predictions from perturbative or lattice QCD (for a review see \cite{Agashe:2014kda}). These different extractions have different levels of sensitivity depending on the order at which the $\alpha_S$ expansion of the observable starts. This order is $\alpha_S^0$ for extractions based on the $R$ ratio. 3-jet rates and event shapes at $e^+e^-$ use observables whose expansion starts at order $\alpha_S^1$. At hadron colliders the ratio of three to two jet production \cite{Chatrchyan:2013txa} and the transfer energy-energy correlation \cite{Aaboud:2017fml} starts at order $\alpha_S^1$, inclusive jet cross section~\cite{Khachatryan:2014waa} starts at order $\alpha_S^2$. The five-jet production rates at LEP \cite{Frederix:2010ne}, the heavy quarkonia hadronic decay width \cite{Brambilla:2007cz} and the 3-jet inclusive observables~\cite{CMS:2014mna} are the observables with the highest sensitivity used so far, their expansion starts at order $\alpha_S^3$. Typically the increased sensitivity comes at a cost, as observables with a lower sensitivity to $\alpha_S$ can be measured more precisely than those with a higher dependency. In this work we present an extraction of $\alpha_S$ using $Z+2,3,4\;\rm jets$ differential cross section measurements from the ATLAS collaboration \cite{ATLASZJets} at a centre of mass energy of $7\;\rm TeV$, comparing them to NLO predictions from \BHS{} \cite{Z4} which start at order $\alpha_S^2$, $\alpha_S^3$ and $\alpha_S^4$ respectively. The increased sensitivity of the higher multiplicity observables partly compensates the larger experimental and theoretical uncertainties in such a way that the three different multiplicities yield comparable degrees of precision for the $\alpha_S$ extraction. We combine the three multiplicities to obtain a final value for $\amz$.   

\section{Extraction procedure}

To obtain our $\alpha_S$ value we compare theoretical predictions obtained from the \BHS{} collaboration~\cite{Z4} with ATLAS data. We minimise the $\chi^2$ function 
\[\chi^2(\alpha_S(M_Z))=\left(y_{t}(\alpha_s(M_Z))-y_{d}\right)^TC^{-1}\left(y_{t}(\alpha_s(M_Z))-y_{d}\right),\]
where $y_t$ are the predictions from theory and $y_d$ are the experimental values. The covariance matrix $C$ is given by
\[C=C_{exp}+C_{pdf}+C_{theory}\;,\]
where $C_{exp}$ is the experimental error covariance matrix, described in section~\ref{sec:exp} and $C_{pdf}$ and $C_{theory}$ are the PDF and theory uncertainty covariance matrices, which we describe in detail in section~\ref{sec:theory}. Our best fit value $\alpha_0$ for $\alpha_S(M_Z)$ is the value that minimises $\chi^2$ and the 1-$\sigma$ interval is given by the values $\alpha_\pm$ of $\amz$ corresponding to $\chi^2(\alpha_\pm)=\chi^2(\alpha_0)+1$. 

To obtain the values $y_t(\alpha_S(M_Z)$ we need to perform a consistent calculation of the theoretical prediction using the same value of $\alpha_S(M_Z)$ in the hard matrix elements as the one used to fit the PDFs. This is possible since many PDF fitting groups provide dedicated fits performed with a range of value of $\alpha_S(M_Z)$. This gives us a discrete set of values for $\chi^2(\alpha_S)$, in order to obtain the precise values of the minimum and 1-$\sigma$ interval we fit a cubic polynomial to the discrete points
and use this fit to determine the minimum $\chi^2$ and the $1-\sigma$ interval. In this work we consider the PDF sets  CT10nlo~\cite{Lai:2010vv}, CT14nlo~\cite{Dulat:2015mca}, MSTW~\cite{Martin:2009iq}, MMHT~\cite{Harland-Lang:2014zoa,Harland-Lang:2015nxa}, NNPDF2.3 ~\cite{NNPDF23} and NNPDF3.0~\cite{NNPDF3}. In the ABM ~\cite{Alekhin:2012ig} PDF set the correlation between the value of $\amz$ and the parameters of the PDFs is stronger than in the other PDF set. As a consequence the $\chi^2$ dependence on $\amz$ is much weaker and does not allow for its determination in our fit.

\subsection{Theoretical prediction}\label{sec:theory}

For the theoretical prediction we used the results of Ref.~\cite{Z4}. In order to perform the extraction procedure and assess uncertainties, we need to re-evaluate the same NLO calculation many times with small modifications. We need to evaluate the prediction for a) different values of the renormalisation and factorisation scales, b) different PDF sets, c) each replica or error set within each PDF set and d) for each value of $\amz$ provided by the PDF set. This type of repetitive calculation with only minor modifications in the PDF and scale setting was one of the motivations behind the development of the n-Tuples format for NLO calculations~\cite{Bern:2013zja}. The other motivation was to allow for the flexibility of defining new observables after the calculation was performed. In our case we do not require this flexibility given that we have settled on the histograms we want use, so we can optimise the amount of recalculation needed by using \fastNLO{}~\cite{fastNLO} tables. We used the public n-Tuples provided by the \BHS{} collaboration for Z+jets~\cite{Z4} to create \fastNLO{} grids allowing the fast re-evaluation of a fixed set of histograms for a different PDF set and different values of the factorisation and renormalisation scales. 

Due to the finite amount of statistics available in the n-Tuples the theoretical predictions have a statistical error. The corresponding covariance matrix can be computed in parallel to the generation of the \fastNLO{} grids. Since the \fastNLO{} library does not report statistical integration errors an alternative method has to be devised to obtain the statistical covariance matrix while avoiding the need to run a full \ntuple{} analysis for each scales and PDF combination. Our strategy is to calculate the statistical covariance matrix for one reference PDF for each scale combination, and then rescale the covariance matrix entries for the other members of the set by the ratio of the bin value for the actual $\amz$ and the reference value:
\[C_{ij}(\amz)\simeq C_{ij}^{ref}\frac{h_i(\amz)}{h_i(\alpha_S^{ref})}\frac{h_j(\amz)}{h_j(\alpha_S^{ref})}\;.\]
This approach assumes that the relative correlation between bins is similar between predictions for different values of $\amz$. We found this assumption to be true at the level of a few percent for the central choice of scale and assumed it to be valid to the same level of accuracy for the other scale choices. Since the statistical uncertainty of the theoretical prediction is not the dominant term in the $\chi^2$ this approximation is well justified. 

The PDF uncertainty is obtained from the PDF error sets using the LHAPDF library \cite{LHAPDF}. We evaluated the NLO prediction for each member of the error set and evaluated the covariance matrices according to the prescriptions of Ref.~\cite{Watt:2012tq} to obtain symmetric errors for the MSTW~\cite{Martin:2009iq}, MMHT~\cite{Harland-Lang:2014zoa,Harland-Lang:2015nxa}, CT10nlo~\cite{Lai:2010vv} and CT14nlo~\cite{Dulat:2015mca} PDFs. 
The covariance matrices of all PDF fits have been rescaled to correspond to a 68\% confidence level if necessary. The covariance matrix for NNPDF2.3 ~\cite{NNPDF23} and NNPDF3.0~\cite{NNPDF3} are obtained statistically from the set of 100 replica.    

\subsection{Scale uncertainty}\label{sec:scale}
The NLO predictions have been carried out using a factorisation and renormalisation scale $\mu_0$ defined in terms of the the jet transverse momenta $p_T^i$ and the mass $M_Z$ of the $Z$ boson
\begin{equation}\label{eq:central}
  \mu_0=\hat{H'}_T/2\;,
\quad
\hat{H'}_T=\sum\limits_i p_T^i+E_T^Z\;,
\quad
E_T^Z=\sqrt{M_Z^2+\left(p_T^Z\right)^2}\;,
\end{equation}
where the sum runs over all partons in the final state. To account for the scale uncertainty we employ two different methods. The first method is to repeat the extraction using predictions obtained with factorisation and renormalisation scales modified from the central scale from Eq.~(\ref{eq:central}) by factors $f_\mu^F,f_\mu^R=1/2,1,2$. The scale uncertainty is taken to be the envelope of the result obtained from all pairs where the factors $m_\mu^F$ and $f_\mu^R$ differ by at most a factor of 2.  

\begin{figure}
  \begin{center}
    \includegraphics[scale=0.4]{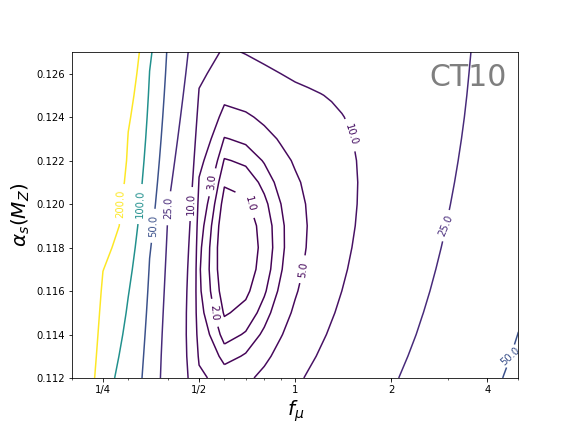}
    \includegraphics[scale=0.4]{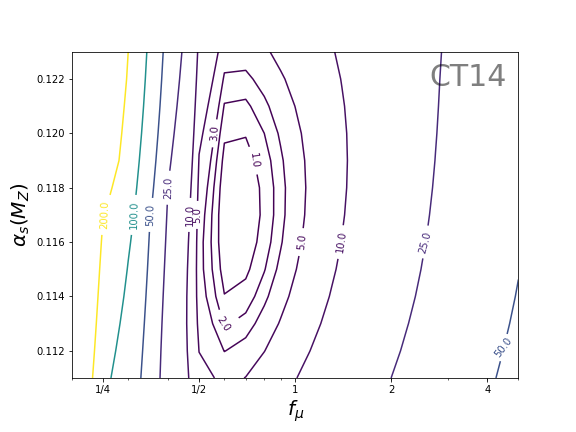}
    \includegraphics[scale=0.4]{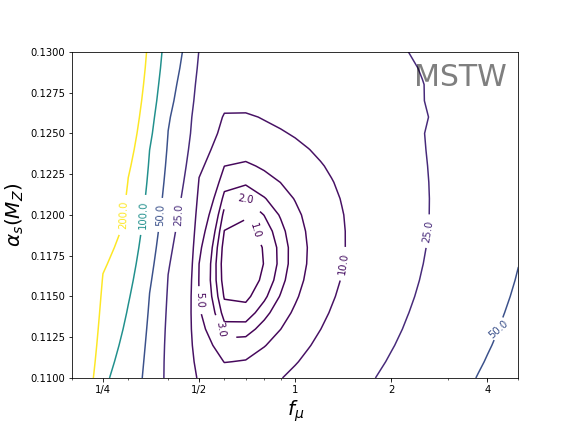}
    \includegraphics[scale=0.4]{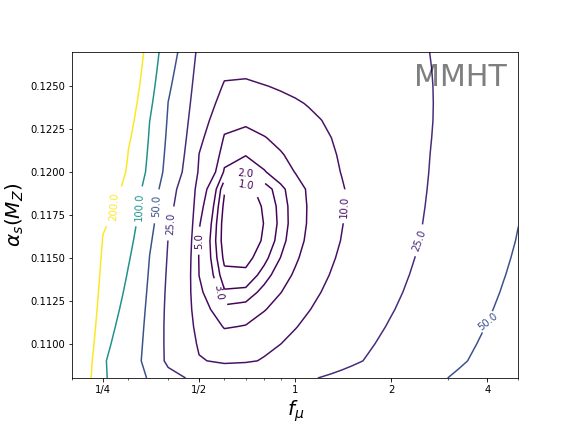}
    \includegraphics[scale=0.4]{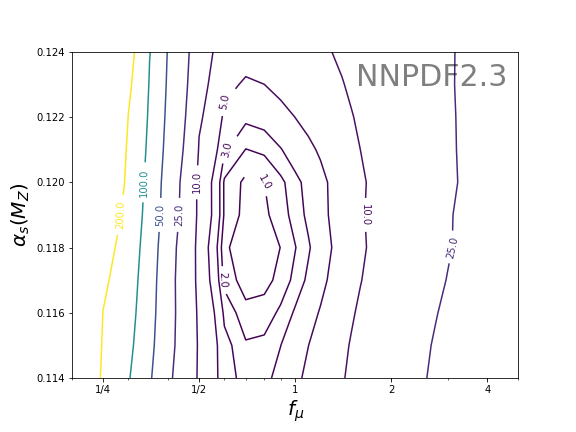}
    \includegraphics[scale=0.4]{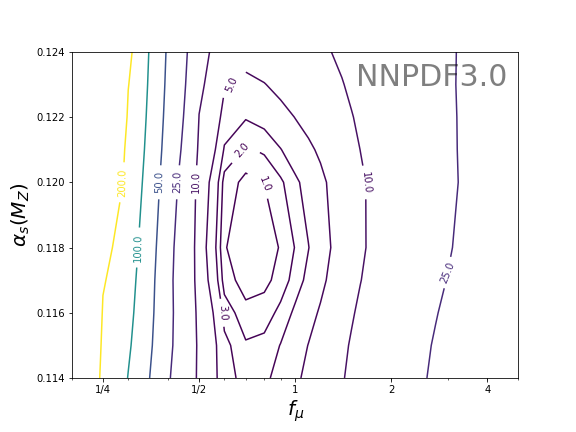}
    \end{center}
  \caption{ $\chi^2$ in the $\alpha_S(M_Z)$-$f_\mu$ two-dimensional plane for all the PDF sets considered. }\label{fig:scale2Dall}
  \end{figure}

In the second method we vary the factorisation and renormalisation scales by a common factor $f_\mu=f_\mu^F=f_\mu^R$ and treat the value of this factor as a nuisance parameter for the fit. To do so we calculate the value of $\chi^2$ for many different values of $f_\mu$ and define the profile $\chi^2$:
\[\hat \chi^2(\amz) = \min\limits_{f_\mu}\chi^2(\amz,f_\mu)\]
and then minimise this function $\hat \chi^2(\amz)$ as a function of $\amz$ to obtain the best fit $\amz$ and 1-$\sigma$ uncertainty interval. Figure~\ref{fig:scale2Dall} shows the $\chi^2$ distributions as a function of $f_\mu$ and $\amz$ for each PDF set. 

Uncertainty intervals obtained in this way account for both experimental, PDF and theoretical error sources, and also for our choice of $f_\mu$. The main advantage of this method is that it does not rely on the somewhat arbitrary values $f_\mu=1/2$ and $f_\mu=2$ used in the traditional approach. As can be seen from Figure~\ref{fig:scale2Dall}, the fit seems to favour slightly smaller scales than the one used for the central scale.


\subsection{Non-perturbative corrections}
Predictions provided by the \BHS{} \ntuples{} are at parton level. In order to correct for hadronisation and underlying event effects we corrected the partonic cross section using the same corrections as used in the experimental comparison to the NLO prediction~\cite{ATLASZJets}. These non-perturbative corrections are estimated by comparing simulated samples generated using {\sc ALPGEN}~\cite{Mangano:2002ea} with and without a fragmentation and underlying event model. 

In order to assess the uncertainty on these corrections two independent models are used for the non-perturbative modelling: the first set of corrections use Herwig+JIMMY~\cite{Corcella:2000bw,Butterworth:1996zw} using the AUET2-CTEQ61L tune~\cite{ATL-PHYS-PUB-2011-008}, the second uses PYTHIA~\cite{Sjostrand:2006za} with the {\sc PERUGIA2011C} tune~\cite{Skands:2010ak}.
In both cases the correction factors have statistical uncertainty due to the size of the simulated samples used to derive them. This uncertainty is added to the theoretical covariance matrix.
As for the scale uncertainty, we use two different methods to estimate the impact of the non-perturbative corrections on our extraction. In the standard method we use the average of the correction factors for each bin to obtain the central prediction, and use the individual correction factors to estimate the uncertainty band. In a more flexible method we combine the two correction factors according to
\begin{equation}\label{eq:mixing}
\delta_{NP}=\lambda \delta_{NP}^{Herwig} + (1-\lambda)\delta_{NP}^{Pythia}.
\end{equation}
The central value of the standard method described above corresponds to $\lambda=1/2$ and the band is given by the values $\lambda=0,1$. In the second method we treat $\lambda$ as a nuisance parameter.
In the tail of the rapidity distributions the NLO description is not expected to be very accurate, which can be seen in an increase of the corrections described above from a few percent to over 10\%. To limit the impact of this corner of phase space to our extraction we combined the last bins of the rapidity distributions into one bin for each multiplicity in such a way that the correction factor for the resulting bin does not exceed 10\%.  

\subsection{Experimental data}\label{sec:exp}
The experimental values we used in this extraction were obtained from measurements of jets produced in association with a Z boson in proton-proton collisions at a centre of mass energy $\sqrt{s}=7\;\mbox{TeV}$. The data corresponds to an integrated luminosity of $4.6\;\rm{fb^{-1}}$ collected by the ATLAS detector. The $Z$ bosons were selected in the electron and muon pair decay channels, the jets were selected with a transverse momentum cut of  $p_T > 30 \rm{GeV}$ and a rapidity cut $|y| < 4.4$. 

For our $\alpha_S$ extraction we use the results presented in~\cite{Z7TeVAtlas} and available from HepData \cite{doi} for the rapidity and transverse momentum distribution of the $n$-th jet in $Z+n\;\rm jets$ events. We corrected the results for the updated total luminosity reported in Ref.~\cite{Aad:2013ucp}. The experimental uncertainties can be separated into three categories: the statistical error, the systematic uncertainty and the luminosity uncertainty. The authors used the procedure described in \cite{Aad:2011he,ATLAS-CONF-2013-004} to separate the correlated uncertainties into a set of independent fully correlated uncertainties which we used to calculate the full experimental covariance matrix. 
\section{Results}
  \begin{figure}
  \begin{center}
    \includegraphics[scale=0.52]{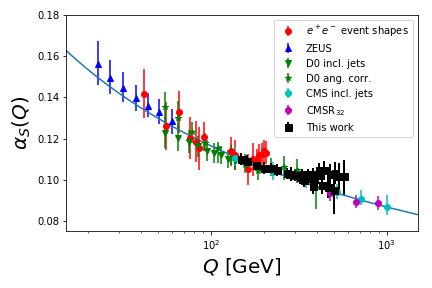}
    \includegraphics[scale=0.52]{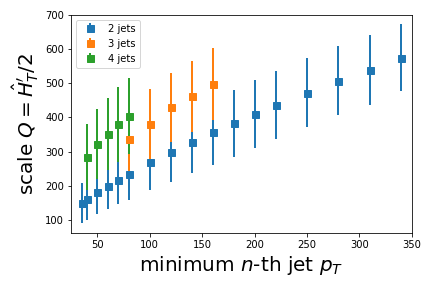}
    \end{center}
  \caption{Left panel: Measurement of the strong coupling constant as a function of $Q^2$. Right panel: Expectation value of  $Q^2$ as a function of the minimum jet transverse momentum considered for the fit. The PDF set MSTW2008~\cite{MSTW2008} was used for these two figures.}\label{fig:alphasOfQ}
  \end{figure}
The left panel of Figure~\ref{fig:alphasOfQ} shows the scale dependence of $\alpha_S(Q)$. The different points on the graph represent the value of $\alpha_S$ obtained by restricting the fit to the transverse momentum of the second, third and fourth jets to the subsets of bins above a given threshold. The value of the scale $Q$ assigned to the fit value is the expectation value of the scale used for the NLO calculation ($\hat{H}_T'/2$ as defined in ref.~\cite{W5j}) for this subset of bins. These scales are shown in the right panel of Figure~\ref{fig:alphasOfQ} for each multiplicity as a function of the minimum transverse momentum in the bin subset. The error bars represent the variance of scale when restricted to the bins above the minimum transverse momentum. The values shown in Fig.~\ref{fig:alphasOfQ} are obtained using the CT10nlo PDF set, the values obtained with other PDF sets are very similar. The values that fall somewhat above the theoretical curve correspond to the highest values of the second jet pt $p_{2,min}^T$. In this very restricted phase-space the three- and four-jets contributions are significant so that the NLO 2-jet calculation underestimates the cross section, resulting in a higher $\amz$ value.

\begin{figure}
  \begin{center}
    \includegraphics[scale=0.5]{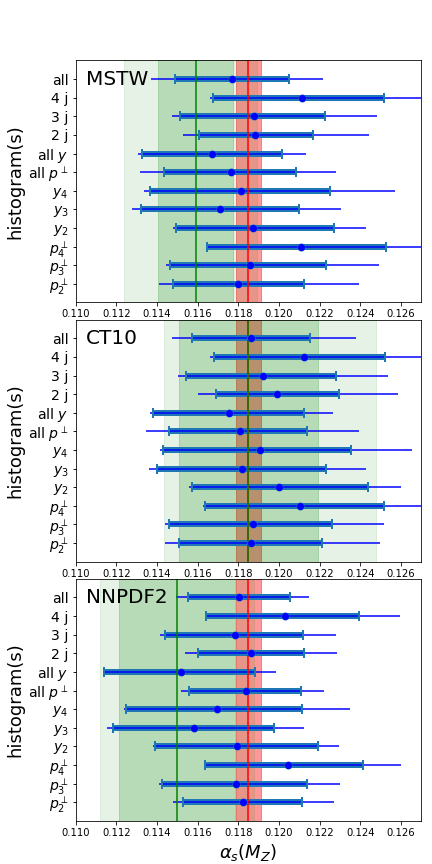}
    \includegraphics[scale=0.5]{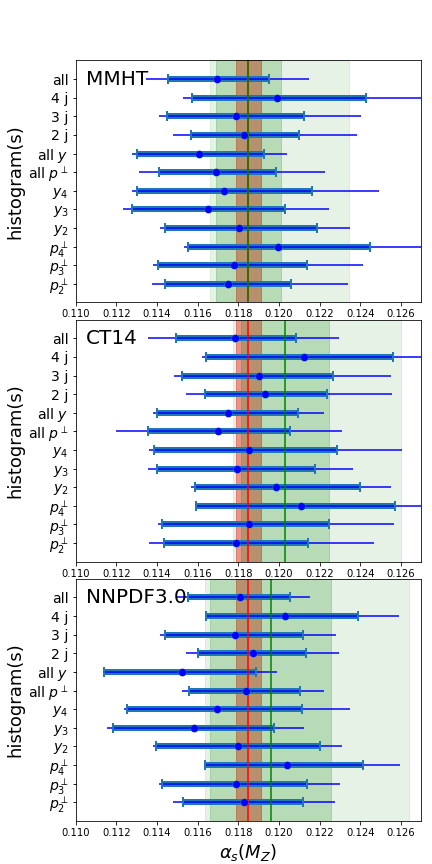}
    \end{center}
  \caption{ Strong coupling values for different PDF sets for each considered histogram and combinations. The green bands on the left-hand plots are the result from \cite{Khachatryan:2014waa}\footnote{The NNPDF results from \cite{Khachatryan:2014waa} used version 2.1 while we used version 2.3 in this work.} and the green bands on the right-hand side show the results from \cite{Aaboud:2017fml}. The darker regions represent the uncertainties without the scale variation and the lighter regions show the total uncertainties including the scale variation. The red band is the world average~\cite{PDG}. The boundaries of the thick part of the error bars represent the values of $\alpha_s(M_Z)$ for which the $\chi^2-\chi^2_{min}=1$. These error estimates do not include scale variation. The thin error bars show the uncertainty including the scale uncertainties.}\label{fig:all}
  \end{figure}

The left-hand column of plots in Figure~\ref{fig:all} shows the best fit results for the PDF sets MSTW2008~\cite{MSTW2008}, CT10~\cite{CT10} and NNPDF~2.3~\cite{NNPDF23}. These sets were chosen to facilitate the comparison with the results obtained in ref.~\cite{Khachatryan:2014waa}. 
The right-hand column of plots in Figure~\ref{fig:all} shows the best fit results for the PDF sets MMHT~\cite{Harland-Lang:2014zoa}, CT14~\cite{Dulat:2015mca} and NNPDF~3.0~\cite{NNPDF3}. The results for this set of PDFs are compared with the results published in Ref.~\cite{Aaboud:2017fml}.
The results are given for fits to the following combinations of observables:
\begin{itemize}
\item each distributions separately,
\item combination of the transverse momentum and rapidity for each multiplicity,
\item the combination of all three transverse momentum distributions,
\item the combination of all the rapidity distributions, 
\item the combination of all histograms.
\end{itemize}
A few patterns emerge from Fig.~\ref{fig:all}: a) the fit to the rapidity histograms favours smaller values of $\amz$ while fits to the rapidity distributions prefer larger values of $\amz$, b) fits for the lower multiplicities tend to yield only moderately more accurate results than higher multiplicity ones, c) the covariance matrix for the rapidity distributions displays a large correlation, causing their combination to yield a value more extreme than any individual result.  

\begin{figure}
  \begin{center}
    \includegraphics[scale=0.6]{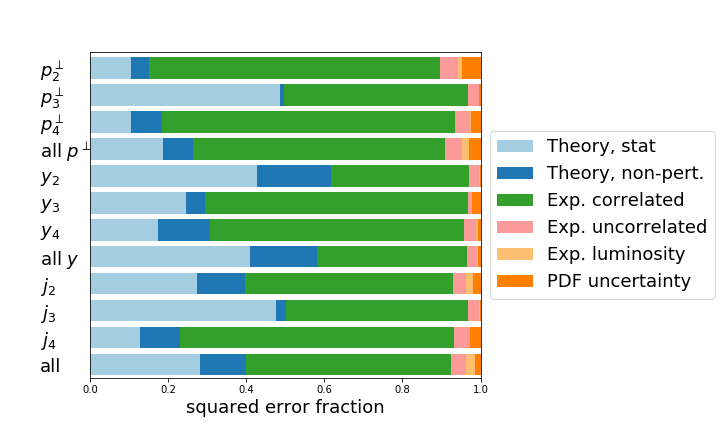}
    \end{center}
  \caption{ Share of the uncertainty by error source according to Eq.~\ref{eq:share}. The results are for the PDF set CT14.}\label{fig:share}
  \end{figure}

To estimate the share of the uncertainties due to each error source we use the quantities
  \begin{equation}\label{eq:share}
    \chi_s^2=\left(y_{t}(\alpha_s(M_Z))-y_{d}\right)^TC_{tot}^{-1}C_sC_{tot}^{-1}\left(y_{t}(\alpha_s(M_Z))-y_{d}\right)
    \end{equation}
defined for each error source covariance matrix $C_s$. The $\chi_s^2$ sum up to the total $\chi^2$. We assign each error source a fraction $\chi^2_s/\chi^2$ of the total uncertainty. In the limit where all errors are fully uncorrelated this procedure is equivalent to summing the errors in quadrature. 
  Figure~\ref{fig:share} shows an example of how the uncertainty is shared between the error sources for CT14. The figure shows the share for each individual distribution, for the combination of all transverse momentum distributions and rapidities, for the multiplicity combination and for the full combination.
We can see that the dominant share of the uncertainty comes from the experimental uncertainties. In principle the uncertainty could be reduced by increasing the statistical accuracy of the theory prediction and the understanding of the non-perturbative corrections. 

Table~\ref{tab:result} shows the result for the best fit $\amz$ for a list of PDF sets. The uncertainties in this table do not include scale variation and non-perturbative correction uncertainties. The theory and experimental uncertainties are approximately of the same size while the PDF uncertainty is almost one order of magnitude smaller. The $\chi^2$ per degree of freedom in slightly below 1 and is very similar across PDF sets. 

\begin{table}
     \caption{Results for the coupling constant extraction with uncertainties.\label{tab:result}}
     \begin{ruledtabular}
     \begin{tabular}{ccccc}

    PDF set   &   $\alpha_s(M_Z)$   & uncertainties detail & $\chi^2/ndof$  \\\hline
    
            CT10nlo
            &
            $0.1186 {}^{+0.0029 }_{ - 0.0029 }$
            &
            ${}^{+0.0018  }_{  -0.0018 }\mbox{(theory)}
             {}^{+0.0022     }_{  -0.0023    }\mbox{(exp)}
             {}^{+0.00035     }_{  -0.00036    }\mbox{(pdf)}
            $
            &
            $ 45.4022 / 60 $
            \\
        
            MSTW2008nlo68cl
            &
            $0.1177 {}^{+0.0028 }_{ - 0.0028 }$
            &
            ${}^{+0.0017  }_{  -0.0017 }\mbox{(theory)}
             {}^{+0.0022     }_{  -0.0022    }\mbox{(exp)}
             {}^{+0.00034     }_{  -0.00034    }\mbox{(pdf)}
            $
            &
            $ 48.4813 / 60 $
            \\
        
            NNPDF2.3\_nlo\_as\_0118
            &
            $0.1180 {}^{+0.0025 }_{ - 0.0025 }$
            &
            ${}^{+0.0016  }_{  -0.0016 }\mbox{(theory)}
             {}^{+0.0019     }_{  -0.0020    }\mbox{(exp)}
             {}^{+0.00022     }_{  -0.00022    }\mbox{(pdf)}
            $
            &
            $ 46.5188 / 60 $
            \\
        
            CT14nlo
            &
            $0.1178 {}^{+0.0030 }_{ - 0.0029 }$
            &
            ${}^{+0.0019  }_{  -0.0018 }\mbox{(theory)}
             {}^{+0.0023     }_{  -0.0022    }\mbox{(exp)}
             {}^{+0.00042     }_{  -0.00041    }\mbox{(pdf)}
            $
            &
            $ 46.7203 / 60 $
            \\
        
            MMHT2014nlo
            &
            $0.1169 {}^{+0.0026 }_{ - 0.0024 }$
            &
            ${}^{+0.0016  }_{  -0.0015 }\mbox{(theory)}
             {}^{+0.0020     }_{  -0.0019    }\mbox{(exp)}
             {}^{+0.00032     }_{  -0.00029    }\mbox{(pdf)}
            $
            &
            $ 47.7737 / 60 $
            \\
        
            NNPDF3.0\_nlo\_as\_0118
            &
            $0.1181 {}^{+0.0025 }_{ - 0.0026 }$
            &
            ${}^{+0.0016  }_{  -0.0016 }\mbox{(theory)}
             {}^{+0.0019     }_{  -0.0020    }\mbox{(exp)}
             {}^{+0.00025     }_{  -0.00025    }\mbox{(pdf)}
            $
            &
            $ 46.3288 / 60 $
            \\

\end{tabular}
         \end{ruledtabular}
         \end{table}

Table~\ref{tab:result_scale} shows scale variation uncertainty estimates using the two methods outlined in section~\ref{sec:scale}. The second and third columns vary the factorisation and renormalisation scales by factors of $1/2$, $1$ or $2$. The second column shows the resulting uncertainty if the same factor is chosen for both scales, and the third column shows the uncertainty resulting from choosing any two factors differing by at most a factor of two. Comparing these two columns we can see that the correlated variation covers most of the range of the uncorrelated variation.

While the 1-$\sigma$ intervals from the $\chi^2$ fit are essentially symmetric the uncertainties due to the first scale variation method are asymmetric, with the upward fluctuation generally much larger than the downward one. Using the nuisance parameter approach to scale variation we get roughly symmetric uncertainty estimates and higher best fit values. The uncertainty intervals are smaller for CT10 but larger for all other PDF sets. This approach gives more symmetric error intervals than the standard variation, as in the case of NNPDF2.3 case where the standard variation gave a very small lower variation in the standard approach.

\begin{table}
     \caption{Scale uncertainties for the $\amz$ extraction. The first column repeats the result of Table~\ref{tab:result} for the theoretical, experimental and PDF uncertainties from the fit. The second and third columns show the uncertainties resulting from varying the scale by either correlated or uncorrelated factors. The fourth column shows the result of the fit where the scale factor is treated as a nuisance parameter.\label{tab:result_scale}}
     \begin{ruledtabular}
     \begin{tabular}{cccccc}

    PDF set   &   $\alpha_s(M_Z)$   &   scale corr. & scale uncorr.  & $f_\mu$ as nuisance param \\\hline
    
            CT10nlo
            &
            $0.1186^{+0.0029 }_{ - 0.0029 }$
            &
            ${} ^{+0.0039 }_{ - 0.0018 }$
            &
            ${} ^{+0.0039 }_{ - 0.0018 }$
            &
            $0.1202^{+0.0025 }_{ - 0.0028 }$
            \\
        
            MSTW2008nlo68cl
            &
            $0.1177^{+0.0028 }_{ - 0.0028 }$
            &
            ${} ^{+0.0023 }_{ - 0.0021 }$
            &
            ${} ^{+0.0027 }_{ - 0.0021 }$
            &
            $0.1195^{+0.0028 }_{ - 0.0028 }$
            \\
        
            NNPDF2.3\_nlo\_as\_0118
            &
            $0.1180^{+0.0025 }_{ - 0.0025 }$
            &
            ${} ^{+0.0017 }_{ - 0.0006 }$
            &
            ${} ^{+0.0017 }_{ - 0.0006 }$
            &
            $0.1196^{+0.0025 }_{ - 0.0025 }$
            \\
        
            CT14nlo
            &
            $0.1178^{+0.0030 }_{ - 0.0029 }$
            &
            ${} ^{+0.0031 }_{ - 0.0025 }$
            &
            ${} ^{+0.0034 }_{ - 0.0025 }$
            &
            $0.1196^{+0.0031 }_{ - 0.0030 }$
            \\
        
            MMHT2014nlo
            &
            $0.1169^{+0.0026 }_{ - 0.0024 }$
            &
            ${} ^{+0.0027 }_{ - 0.0019 }$
            &
            ${} ^{+0.0030 }_{ - 0.0019 }$
            &
            $0.1184^{+0.0027 }_{ - 0.0025 }$
            \\
        
            NNPDF3.0\_nlo\_as\_0118
            &
            $0.1181^{+0.0025 }_{ - 0.0026 }$
            &
            ${} ^{+0.0017 }_{ - 0.0003 }$
            &
            ${} ^{+0.0017 }_{ - 0.0003 }$
            &
            $0.1196^{+0.0025 }_{ - 0.0025 }$
            \\

\end{tabular}
         \end{ruledtabular}
         \end{table}

Table~\ref{tab:result_corr} collects the results for the assessment of the non-perturbative corrections. Using the correction factors calculated using ALPGEN+Herwig leads to higher values of $\amz$ than when using ALPGEN+Pythia. The difference between the results obtained with the two sets of program is quite large and is commensurate with the other uncertainties affecting the fit. Such a large disagreement in the values of non-perturbative corrections is not uncommon, see for example \cite{Aaboud:2017dvo,Aaboud:2017fml}. 

\begin{table}
     \caption{Non-perturbative uncertainty for the $\amz$ extraction. The results in first column are obtained by correcting each bin with the average of the correction factors obtained with ALPGEN+Herwig and ALPGEN+Pythia. The second and third column are the results obtained using only the individual programs. The last column is the uncertainty, taken as the difference between the central values and the second and third columns.  \label{tab:result_corr}}
     \begin{ruledtabular}
     \begin{tabular}{cccccc}

    PDF set   &   Central value   & ALPGEN+Herwig  & ALPGEN+Pythia &  uncertainty  \\\hline
    
            CT10nlo
            &
            $0.1186$
            &
            $0.1205$
            &
            $0.1168$
            &
            ${}^{ +0.0018 }_{ -0.0018 }$
            \\
        
            MSTW2008nlo68cl
            &
            $0.1177$
            &
            $0.1199$
            &
            $0.1158$
            &
            ${}^{ +0.0022 }_{ -0.0019 }$
            \\
        
            NNPDF2.3\_nlo\_as\_0118
            &
            $0.1180$
            &
            $0.1198$
            &
            $0.1164$
            &
            ${}^{ +0.0017 }_{ -0.0017 }$
            \\
        
            CT14nlo
            &
            $0.1178$
            &
            $0.1202$
            &
            $0.1159$
            &
            ${}^{ +0.0023 }_{ -0.0019 }$
            \\
        
            MMHT2014nlo
            &
            $0.1169$
            &
            $0.1192$
            &
            $0.1154$
            &
            ${}^{ +0.0023 }_{ -0.0016 }$
            \\
        
            NNPDF3.0\_nlo\_as\_0118
            &
            $0.1181$
            &
            $0.1197$
            &
            $0.1164$
            &
            ${}^{ +0.0017 }_{ -0.0017 }$
            \\

\end{tabular}
         \end{ruledtabular}
         \end{table}

The results of our extraction are summarised in Table~\ref{tab:result_summary}. The first columns show the results using the standard approach to estimating the scale and non-perturbative uncertainties, while the last column shows the result of the approach where both the scale factor and the mixing parameter $\lambda$ in Eq.~\ref{eq:mixing} are treated as nuisance parameters in the fit. Treating the scale factor and the non-perturbative mixing parameter $\lambda$ as nuisance parameter leads to a smaller value of $\amz$ for all PDF sets except for the NNPDF sets where the best fit value increases. The uncertainty intervals are smaller for NNPDF2.3, NNPDF3, CT14, and CT10 but larger for MSTW and MMHT.   
\begin{table}
     \caption{Summary of the extraction uncertainty. The second column lists the best fit value of the fit with the associated uncertainties. The third and fourth columns show the estimates of the non-perturbative and scale uncertainties using the standard method. The last column shows the result of a fit where both the scale factor and the relative weight of the correction factors is treated as a nuisance parameter.\label{tab:result_summary}}
     \begin{ruledtabular}
     \begin{tabular}{ccccccc}

     PDF set   &    total (standard) &  total (nuisance)  \\\hline
     
             CT10nlo
             &
            $0.1186
            ^{+0.0029 }_{ - 0.0029 }(\mbox{fit})
            {}^{ +0.0018 }_{ -0.0018 }(\mbox{NP})
            {}^{ +0.0039 }_{ - 0.0018 }(\mbox{scale})
            =
            0.1186^{+0.0052 }_{ - 0.0039 }$
             &
            $0.1177^{+0.0037 }_{ - 0.0041 }$
             \\
         
             MSTW2008nlo68cl
             &
            $0.1177
            ^{+0.0028 }_{ - 0.0028 }(\mbox{fit})
            {}^{ +0.0022 }_{ -0.0019 }(\mbox{NP})
            {}^{ +0.0027 }_{ - 0.0021 }(\mbox{scale})
            =
            0.1177^{+0.0045 }_{ - 0.0040 }$
             &
            $0.1177^{+0.0036 }_{ - 0.0038 }$
             \\
         
             NNPDF2.3\_nlo\_as\_0118
             &
            $0.1180
            ^{+0.0025 }_{ - 0.0025 }(\mbox{fit})
            {}^{ +0.0017 }_{ -0.0017 }(\mbox{NP})
            {}^{ +0.0017 }_{ - 0.0006 }(\mbox{scale})
            =
            0.1180^{+0.0035 }_{ - 0.0031 }$
             &
            $0.1197^{+0.0025 }_{ - 0.0031 }$
             \\
         
             CT14nlo
             &
            $0.1178
            ^{+0.0030 }_{ - 0.0029 }(\mbox{fit})
            {}^{ +0.0023 }_{ -0.0019 }(\mbox{NP})
            {}^{ +0.0034 }_{ - 0.0025 }(\mbox{scale})
            =
            0.1178^{+0.0051 }_{ - 0.0043 }$
             &
            $0.1160^{+0.0044 }_{ - 0.0037 }$
             \\
         
             MMHT2014nlo
             &
            $0.1169
            ^{+0.0026 }_{ - 0.0024 }(\mbox{fit})
            {}^{ +0.0023 }_{ -0.0016 }(\mbox{NP})
            {}^{ +0.0030 }_{ - 0.0019 }(\mbox{scale})
            =
            0.1169^{+0.0046 }_{ - 0.0034 }$
             &
            $0.1166^{+0.0030 }_{ - 0.0028 }$
             \\
         
             NNPDF3.0\_nlo\_as\_0118
             &
            $0.1181
            ^{+0.0025 }_{ - 0.0026 }(\mbox{fit})
            {}^{ +0.0017 }_{ -0.0017 }(\mbox{NP})
            {}^{ +0.0017 }_{ - 0.0003 }(\mbox{scale})
            =
            0.1181^{+0.0034 }_{ - 0.0031 }$
             &
            $0.1197^{+0.0026 }_{ - 0.0031 }$
             \\

\end{tabular}
         \end{ruledtabular}
         \end{table}

\begin{figure}[h]
  \begin{center}
    \includegraphics[scale=0.6]{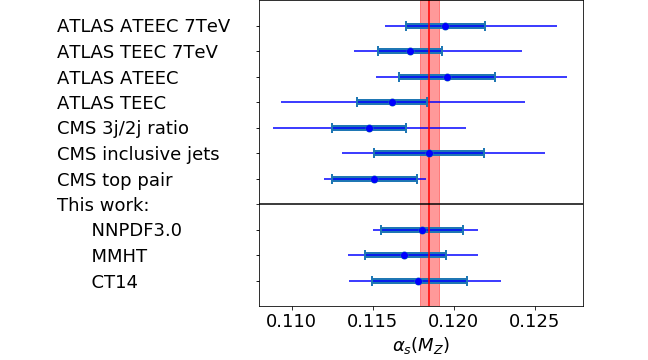}
    \end{center}
  \caption{Comparison of the results presented in this work and other strong coupling extractions using LHC data. The thick part of the error bar includes all errors but the scale uncertainty. The narrow part of the error bar takes that uncertainty into account.}\label{fig:summary}
  \end{figure}

Figure~\ref{fig:summary} shows our results alongside other extractions of $\amz$~\cite{Chatrchyan:2013txa,CMS:2014mna,Khachatryan:2014waa,Khachatryan:2014waa,ATLAS:2015yaa,Chatrchyan:2013haa} using LHC data. 
In general our results have comparable experimental uncertainties. With the exception of the CMS extraction from $t\bar t$ production the result we obtained display a smaller scale uncertainty than the other extractions using LHC data. This observation can be explained by the fact that at NLO the scale uncertainty does not increase with a constant multiplicative factor with each additional power of the coupling constant. This is an important advantage of using high-multiplicity processes to extract a measurement of $\amz$ since adding more experimental data and improving the experimental uncertainties will improve the accuracy of the extraction significantly, while the accuracy of other methods are already limited by the scale contribution to the overall uncertainty. As our final result we choose the value obtained using CT14, using the conventional scale and NP uncertainty estimation method as they result in the most conservative result:
\begin{eqnarray*}
\amz&=&
            {}^{+0.0019  }_{  -0.0018 }\mbox{(theory)}
            {}^{+0.0023     }_{  -0.0022    }\mbox{(exp)}
            {}^{+0.00042     }_{  -0.00041    }\mbox{(pdf)}
            ^{+0.0034 }_{ - 0.0025 }\mbox{(scale)}
            \\&=&
            0.1178^{+0.0030 }_{ - 0.0029 }\mbox{(all but scale)}
            {} ^{+0.0034 }_{ - 0.0025 }\mbox{(scale)}
            \\&=&
            0.1178 ^{+0.0051 }_{ - 0.0043 }
\;.
\end{eqnarray*}
\section{Conclusion}
 We presented an extraction of the strong coupling constant from high multiplicity $Z$+ jets processes. Our most conservative result is $\amz=0.1178^{+0.0051 }_{ - 0.0043 }$, obtained with the CT14 PDF set. Table~\ref{tab:result_summary} and Fig.~\ref{fig:summary} summarise the best fit values and uncertainty estimates for other PDF sets. We used two different methods to assert the uncertainties from the scale variation and the non-perturbative corrections. Both method yield comparable results. The accuracy obtained for the value of $\alpha_S(M_Z)$ is comparable with other NLO determinations at the LHC, but have a smaller scale uncertainty and a larger experimental uncertainty, leading to compatible estimates. 

The results we obtained show the potential of high-multiplicity processes for the extraction of the strong coupling constant: the larger experimental uncertainties are mostly compensated by the steeper dependence on $\amz$. The lower multiplicities have a slightly smaller uncertainties but the higher multiplicity processes still contribute to the reduction of the final uncertainty. Our results highlight another advantage of higher multiplicity processes, as they have a relatively smaller scale uncertainty, they provide complimentary information to other measurements at the LHC for which the scale variation is the major source of uncertainty.  

The lowest hanging fruit to improve on the uncertainty of the results is to improve the statistical precision of the theoretical prediction. The extraction can be improved with the larger statistics of more recent runs at the LHC and a reduction in the systematic errors of the measurement. A better understanding of non-perturbative effects will also improve the accuracy of the extraction significantly. It is reasonable to expect improvements on all these fronts in the future so we can expect large multiplicity processes to provide improved constraints on the value of the strong coupling constant in the future.    

 \section*{Acknowledgements}
We would like to thank Nigel Glover, Klaus Rabbertz and Stefano Forte for useful discussions and we are grateful to Ulla Blumenschein for providing the non-perturbative corrections used in this work before their publication and for her help with the experimental uncertainties.


 \bibliography{alpha}
\bibliographystyle{apsrev4-1}
\end{document}